\documentstyle[12pt]{article}
\begin{document}

\author{S. L. Lyakhovich, A. A. Sharapov and K. M. Shekhter}
\title{MASSIVE SPINNING PARTICLE IN ANY DIMENSION I. INTEGER SPINS}
\date{{\it Physics Department of Tomsk State University, 634050 Tomsk, Russia }}
\maketitle

\begin{abstract}
The Kirillov-Souriau-Kostant construction is applied to derive the classical
and quantum mechanics for the massive spinning particle in arbitrary
dimension.

\end{abstract}


\section{Introduction}

In the seminal work \cite{Wigner} E. Wigner found that the quantum
mechanical description of any elementary particle should rely on
an irreducible representation of the Poincar\'e group. It
remained unclear, however, what was a classical mechanical model for the
spinning particle which should underlie the respective elementary
particle quantum theory.

The decisive break-through in understanding of the question was reached in
the works of Kirillov \cite{Kirillov}, Kostant \cite{Kostant} and Souriau
\cite{Souriau}. The main subject of their approach, usually referred to
as an {\it orbit method}, is coadjoint orbit of the group $G$ for which
the system is `an elementary one'. The orbits are known to be the
homogeneous symplectic manifolds and thereby they can be naturally
associated with the phase spaces of the elementary dynamical systems. In
the case of relativistic symmetry groups (Poincar\'e, de Sitter, anti-de
Sitter, Galilean) these systems can be conceptually treated as the point
particles with spin. For $4D$ Poincar\'e group, the geometric
classification of relativistic particles on the basis of coadjoint orbits
was first performed by Souriau.

The important observation is that, under certain conditions, quantization of
such elementary systems gives rise to unitary irreducible representations
of the underlying symmetry group $G$. Moreover, there is a broad class of
groups, which has one-to-one correspondence between the orbits and the
representations. Thus the orbit method allows to bind together two, at
first sight different, mathematical disciplines: the symplectic geometry
and the theory of group representations.

Despite its generosity, the orbit method gives rise to rather weak
restrictions on the configuration space of the spinning particle.
Ambiguity in its choice in particular explains a number of different
formulations of the $D=4$ massive spinning particle models. The
configuration space is usually chosen as a direct product of the Minkowski
space and some inner manifold describing spinning degrees of freedom. All
the descriptions of the particle with given mass and spin prove to be
equivalent on the level of the physical phase space (which is just a
coadjoint orbit). Thus for a free relativistic particle the choice of the
initial configuration space is inessential. The picture changes
drastically when trying to switch on the interaction with background
fields. If the system possesses extra gauge symmetry in the Minkowski
space besides the world-line diffeomorphisms it could obstruct consistent
coupling to the background. Physically this follows from the fact that the
space points, equivalent at the free level, can be distinguished by the
background field. In terms of constrained dynamics such systems are
characterized by the first class constraints, generating the gauge
transformations which actually involve spatial variables. The space
inhomogeneity introduced by the external field may destroy the first
class constraints algebra. So, to allow consistent coupling with the
arbitrary background, the system may have only one first class constraint
on the space-time variables (mass-shell) which is related to
reparametrization invariance.

Perhaps, another trend should be mentioned in describing the
relativistic spinning particle on a basis of pseudomechanics with the
Grassmann (anticommuting) variables (see e.g. \cite{BM-GT}). This approach
constitutes an interesting alternative. In particular, the quantization of
such systems in a natural way leads to the finite component relativistic
wave equations. Although these models are well-adapted to the
quantization, the possibility of their classical interpretation is less
clear. In our opinion, the Kirillov-Kostant-Souriau theory, which has a
transparent classical mechanical background, provides a perfectly well
behaved description of a phase space for spin of the particle (as well as
isospin \cite{Duval}) in terms of homogeneous K\"ahler manifolds. We would
like to note, however, that the anticommuting variables seem to be
essential for the formulation of supersymmetric particles.

Except for the models of superparticles, which are traditionally considered
in ten dimensions for their close relationship with superstring, most of the
known spinning particle models (with commuting or anti-commuting variables)
are formulated in four-dimensional space-time (See \cite{Frydryszak} and
references therein). However, there is a number of fundamental physical
theories (superstrings, superbranes and etc.) requiring for their consistent
formulation more than four dimensions. In this connection, the study of the
spinning particles seems to be the natural step in understanding of higher
dimensional physics.

By now the history of spinning particles counts more than 70 years going
back to the pioneer work of Frenkel \cite{Frenkel} and during this period
the reasons for their study have changed permanently. In recent years,
the spinning (super)particles were mainly considered as a toy models
which are useful, for instance, for polishing the string theory
quantization methods. At the present time there is a growing interest in
the study of supersymmetric extended objects -- superbranes -- which are
considered as the basic ingredients of a fundamental, yet unknown, theory
unifying all five consistent superstring theories `under the umbrella of
eleven dimensions' \cite{Duff}, \cite {Townsend}. The existing superbrane
models are formulated in various dimensions so that the physical
space-time appears from the Kaluza-Klein compactification of extra
dimensions. In this context the spinning (super)particles models may gain
a new interest as special examples of $0$ -branes. It is tempting to try
to interpret the spinning particle as a $0$ -brane in the higher dimension
and identify the curled dimensions with the spinning degrees of freedom of
the particle.  Or, even more generally, one may consider $p$-brane and try
to obtain a spinning particle model by simultaneous target space and world
volume reduction. The result of such a reduction would be somewhat
different from the one in the scenario of Kaluza-Klein compactification
where the curled out dimensions are usually responsible for isospin rather
than spin.

The main objective of this work is to present the explicit construction
for the general model of massive spinning particle in arbitrary dimensional
space-time.

The paper is organized as follows. In Sec. 2 we derive the first order
action for the model in the spirit of Soiriau approach. To avoid the
ambiguity in the choice of the initial presymplectic manifold (extended
phase space) of the model it is postulated that the space-time dynamics of
the particle corresponds to one-dimensional world-lines. We describe
the parametrization for the phase space of the model by the vector
variables which provides manifestly covariant form for the dynamics.
Here we also obtain the solutions for the classical equations of motion.
In Sec. 3 we pass to the Lagrange formulation of the model and discuss
admissible boundary conditions. We show that gauge invariance of the
quantum theory is consistent with boundary conditions only under
requirement the spin to be quantized. The Sec. 4 is devoted to the
geometric quantization of the model. The space of physical states is shown
to carry an irreducible representation of the Poincar\'e group of
arbitrary (but fixed) integer spin. The physical wave functions turn out
to be in one-to-one correspondence with the irreducible tensor fields on
the Minkowski space. In Sec. 5 the model is generalized to the case of
the minimal coupling to background electromagnetic and gravitational
fields.  We also derive the equation of motion for this case and shortly
discuss their structure. In Conclusion we summarize the results and touch
on the question of the half-integer spin particles.

\section{Phase-space and the Hamiltonian dynamics.}

It is well known that physical phase spaces for the massive spinning
particles in ${\bf R}^{d-1,1}$ can be associated with the regular coadjoint
orbits of the Poincar\'e group. These orbits, denoted as ${\cal O}_{m,{\bf s}
}$, are labeled by $r+1$ numbers $m$ and ${\bf s}=(s_1,\ldots ,s_r)$, where $
m$ corresponds to the mass of the particle while ${\bf s}$ describes the
particle's spin, $r=[(d-1)/2]$ is a rank of space rotation group $SO(d-1)$.
The regularity means that among all the orbits of Poincar\'e group, ${\cal O}
_{m,{\bf s}}$ have the maximal possible dimension, more precisely, $\dim
{\cal O}_{m,{\bf s}}=(d-1)(d+2)/2-r$ . The orbit ${\cal O}_{m,{\bf s}}$,
being the homogeneous symplectic manifold of Poincar\'e group carries an
invariant closed two-form, which we will denote $\omega $.

In order to formulate the classical dynamics of the particle we follow
the Souriau scheme \cite{Souriau}, which provides the most uniform
and elegant treatment for both the space-time and phase-space evolution of
the system, as well as reveals the role of the symmetries underlying the
theory. When applied to our case this scheme can be summarized by the
following double fibration diagram:

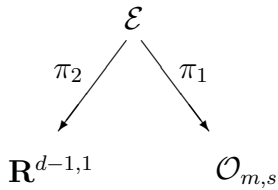
\unitlength=1mm \special{em:linewidth 0.4pt} \linethickness{0.4pt}
\begin{picture}(00.00,35.00)(0,60)
\put(60.00,87.00){\vector(-3,-4){9.00}}
\put(62.00,87.00){\vector(3,-4){9.00}}
\put(61.00,90.00){\makebox(0,0)[cc]{${\cal E}$}}
\put(50.00,70.00){\makebox(0,0)[cc]{${\bf R}^{d-1,1}$}}
\put(76.00,69.50){\makebox(0,0)[cc]{${\cal O}_{m,s}$}}
\put(69.00,82.67){\makebox(0,0)[cc]{$\pi_1$}}
\put(52.33,82.67){\makebox(0,0)[cc]{$\pi_2$}}
\end{picture}

The physical phase space ${\cal O}_{m,{\bf s}}$ of the model is obtained by
a symplectic reduction of the presymplectic manifold ${\cal E}$ equipped by
a closed degenerate two-form $\Omega $ having a constant rank. A generic
fibre $\pi _1^{-1}(a)\subset {\cal E}$ over $a\in {\cal O}_{m,{\bf s}}$ is
associated with the leaf of smooth foliation, generated by $\ker \Omega $,
passing through $a$. In other words, $\pi _1:{\cal E}\rightarrow {\cal O}_{m,
{\bf s}}\simeq {\cal E}/\ker \Omega $ and the presymplectic two-form $\Omega
$ is a pull-back of the symplectic form $\omega $, i.e. $\Omega =\pi
_1^{*}\omega $. The possible space-time evolution of the particle is then
obtained by projecting the fibres $\pi _1^{-1}(a),\ \forall a\in {\cal O}_{m,
{\bf s}}$ into ${\bf R}^{d-1,1}$. Normally, for the case of massive
particles, the image of $\pi _2\pi _1^{-1}:{\cal O}_{m,{\bf s}}\rightarrow
{\bf R}^{d-1,1}$ are (nonparametrized) time-like geodesics in Minkowski
space \cite{Souriau}, but also higher-dimensional surfaces may
occur \cite{Zakrzewski}, \cite{ms-particle}, \cite{uni-model},
\cite{Guillemin}.

As $\Omega $ is closed, there exists (at least locally) a potential
one-form $\vartheta $, so that $\Omega =d\vartheta $. Then the first order
(Hamiltonian) action of the theory can be chosen as
$$
S=\int\limits_\gamma \vartheta
$$
There is an obvious ambiguity in the possible choice of $ {\cal E}$,
although the symmetries of the model, if any, may hint at suitable
manifold. Thus, for the {\it elementary dynamical system} (as in the case
at hand) an appropriate presymplectic manifold is to be looked for among
the homogeneous spaces of the symmetry group for which the system is an
elementary one. This condition, however, still does not specify the
choice of $ {\cal E}$ completely and we postulate an additional
requirement that the space-time projection of the particle's trajectory on
${\cal E}$ is a one-dimensional world line. The physical reasons for this
restriction lie mainly in the observation that nonlocal dynamics of the
particle in ${\bf R} ^{d-1,1}$ seems to allow no self-consistent coupling
to external gauge fields \cite{ms-particle}, \cite{d6-particle}. Finally,
the space ${\cal E}$ must be wide enough to provide nontrivial dynamics
for an arbitrary spin.

The simplest way to satisfy the above conditions is to find the homogeneous
space ${\cal E}$ for which $\dim {\cal E}=\dim {\cal O}_{m,{\bf s}}+1$. This
unambiguously leads to the quotient ${\cal E}\simeq
(Poincar\acute e\ group)/H,$ where $H\subset SO(d-1)\subset SO(d-1,1)$ is
the Cartan subgroup of the group of space rotations, $H\simeq [SO(2)]^r$.
Using the Iwasawa decomposition for the Lorentz group one may readily see
the following global structure of ${\cal E}$:  \begin{equation} \label{b1}
\begin{array}{c}
{\cal E}={\cal E}_{spinless}\times {\cal O}_{{\bf s}}\ , \\  \\
{\cal E}_{spinless}={\bf R}^{d-1,1}\times B\ ,\qquad {\cal O}_{{\bf s}
}=SO(d-1)/[SO(2)]^r
\end{array}
\end{equation}
Here we identify the $d$-dimensional Minkowski space with the subgroup of
Poincar\'e translations and $B$ stands for the set of boosts, which can be
naturally identified with the upper sheet ($p_0>0$) of mass hyperboloid, $
p_Ap^A+m^2=0$, $p$ being $d$-momentum of the particle. Then the first factor
${\cal E}_{spinless}$ is nothing but the presymplectic manifold for the
massive spinless particle equipped with the conventional two-form $
dp_A\wedge dx^A$. The second factor ${\cal O}_{{\bf s}}$ is a regular
coadjoint orbit of $SO(d-1)$, thus it constitutes phase space for the
spin. The direct product structure of (\ref {b1}) implies that the
invariant presymplectic structure $\Omega $ on $ {\cal E}$ can be obtained
as the sum of ones on ${\cal E}_{spinless}$ and $ {\cal O}_{{\bf s}}$.
Note that since $H$ belongs to the stability subgroup of $p$,
the symplectic form on ${\cal O}_{{\bf s}}$ may depend on the points of
mass hyperboloid. The explicitly covariant expression for $\Omega $ is
constructed as follows. Let $V\subset {\bf C}^{dr}$ be an open
domain associated with the set of complex $d\times r$ matrices $Z$
\begin{equation}
\label{b3}Z^t=(z_i^A)^t=\left( \begin{array}{ccccc} z_1^0
& z_1^1 & \cdots  &  & z_1^{d-1} \\ \vdots  & \vdots  &  &  & \vdots  \\
z_r^0 & z_r^1 & \cdots  &  & z_r^{d-1}
\end{array}
\right) \quad ,\qquad rank\ Z=r
\end{equation}

Consider a surface in $V$ extracted by the set of algebraic
equations
\addtocounter{equation}{1}
\label{b4}
$$
pZ=0
\eqno{(3.a)} $$
$$
Z^t\eta Z=0\quad ,\qquad diag\ \eta =(-+\cdots +)
\eqno{(3.b)} $$
The space ${\cal O}_{{\bf s}}$ can be now obtained after
factorization of this surface
by the following equivalence relation:
\begin{equation} \label{b5}Z^{\prime }\sim Z\cdot \Lambda
\end{equation}
where $\Lambda $'s are nondegenerate upper triangular $r\times r$ matrices
\begin{equation}
\label{a3}\Lambda =\left(
\begin{array}{cc}
\ddots  & * \\
0 & \ddots
\end{array}
\right) \quad ,\qquad \det \Lambda \neq 0
\end{equation}
The action of the Lorentz group on ${\cal O}_{{\bf s}}$ is induced by the
left multiplication of $Z$ by $G\subset SO(D-1,1)$ in fundamental
representation
\begin{equation}
Z^{\prime }=G\cdot Z
\end{equation}
In the rest frame, $\stackrel{\circ }{p}=(m,0,\ldots ,0)$, one readily
recognizes in (\ref{b3}-\ref{b5}) the flag manifold construction
for orthogonal groups. In order to assign a K\"ahler structure to ${\cal
O}_{{\bf s}}$, we introduce the positive-definite Hermitian $r\times r$
matrix $D=Z^{\dagger }Z$ and define the set of its principle minors
\begin{equation}
\label{a7}\Delta _k=\left|
\begin{array}{ccc}
D_1^1 & \cdots  & D_1^k \\
\vdots  &  & \vdots  \\
D_k^1 & \cdots  & D_k^k
\end{array}
\right| \quad ,\qquad k=1,\ldots ,r
\end{equation}
Due to the special structure of $\Lambda $-matrices all $\Delta $'s are
invariant under the equivalence transformations (\ref{b5}) up to constant
multiplier and all the possible invariants are exhausted by
the homogeneous polynomials in $\Delta $'s. On this basis the most general
expression for the K\"ahler potential can be written as
\begin{equation}
\label{b7}\Phi =\frac 12\ln (\Delta _1^{s_1}\Delta _2^{s_2}\cdots \Delta
_r^{s_r})
\end{equation}
The parameters $s_i$ in (\ref {b7}) are fixed by the particular choice of
the coadjoint orbit ${\cal O}_{\bf s}$. In what follows they will be
related to the spin of the particle. Moreover, the quantum
self-consistency will restrict $s_i$ to be integers. Transformations
(\ref{b5}) change $\Phi $ by an additive constant
\begin{equation}
\label{a9}\delta _\Lambda
\Phi =\sum_{k=1}^r\ln \left| \Lambda _1^1\Lambda _2^2\cdots \Lambda
_k^k\right| ^{s_k}
\end{equation}
and therefore the K\"ahler two-form
$d*d\Phi $ is well defined on ${\cal O}_{{\bf s}}$.  (Here we introduce
the star operator $*$ , which action on complex one-forms is defined by
the rule: $*(\alpha _Idz^I+\beta _Id \overline{z}^I)=-i(\alpha
_Idz^I-\beta _Id\overline{z}^I)$). Then the presymplectic two-form on
${\cal O}_{m,{\bf s}}$ can be written as
\begin{equation} \label{b6}\Omega
=dp_A\wedge dx^A+d*d\Phi
\end{equation}
According to the general prescription discussed above, the Hamiltonian
action of the model can be chosen as
\begin{equation}
\label{b8}
\begin{array}{c} S=\int \vartheta \ , \\ \\ \vartheta =pdx+*d\Phi
\end{array}
\end{equation}
The variation principle being applied to this
action yields
\begin{equation}
\label{b9}\stackrel{.}{\Gamma }\rfloor
\Omega =0
\end{equation}
where $\Gamma =(x,p,Z,\overline{Z})$.
Since $\ker
\Omega $ is spanned by the only vector field
\begin{equation}
\label{b10}{\bf V}=p^A\frac \partial {\partial x^A}\ ,
\end{equation}
associated with the mass-shell condition $p^2+m^2=0$, eq. (\ref{b9}) is
equivalent to
\begin{equation}
\label{b11}
\begin{array}{c}
\stackrel{.}{x}=\lambda p\quad ,\qquad \stackrel{.}{p}=0 \\  \\
\stackrel{.}{Z}=0\qquad (modulo\ \Lambda -\ transformations) \end{array}
\end{equation}
$\lambda $ being a Lagrange multiplier associated with the reparametrization
invariance. We see that in this theory the particle moves along the
time-like geodesics in $ {\bf R}^{d-1,1}$ whereas its dynamics in
internal space is `frozen'.  Thus the physical phase space of the model
${\cal O}_{m,{\bf s}}\simeq {\cal E} /\ker \Omega $ is isomorphic to the
quotient space $(Poincar\acute e\ group)/H\times T$, where $T$ corresponds
to the time translations.

\section{Lagrangian description, gauge invariance and spin quantization}

The first-order action (\ref{b8}) may be readily brought to the
second-order form (w.r.t. $x$). Proceeding in the spirit of constrained
dynamics, the action (\ref{b8}) is added by the mass-shell condition
$p^2+m^2\approx 0$ and $p$-transversality conditions $pZ\approx 0,\ p\bar
Z\approx 0$ taken with arbitrary Lagrange multipliers. Eliminating then
$p$ and the Lagrange multipliers by their equations of motion, we arrive
at the action functional
\begin{equation}
\label{c1}S=\int d\tau \left\{
m\sqrt{-\stackrel{.}{x}{}^2+2(\stackrel{.}{x}
,z_i)(D^{-1})^{ij}(\stackrel{.}{x},\overline{z}_j)}-i(\stackrel{.}{z}
\!_i^A\partial _A^i\Phi -c.c)\right\}
\end{equation}
The configuration space of the model is the direct product ${\bf R}
^{d-1,1}\times \Sigma $, where $\Sigma$ is the surface in ${\bf C}^{dr}$
extracted by $(3.b)$. The action is obviously Poincar\'e and
reparametrization invariant and possesses, by the construction, the gauge
symmetries associated with the equivalence relation (\ref{b5})
\begin{equation}
\label{c2}Z^{\prime }(\tau )=Z(\tau )\Lambda (\tau )
\end{equation}
These transformations result in the modification of
(\ref{c1}) by a function of the gauge parameters at the integral limits
$$
\delta S=\frac 12\int\limits_{\tau ^{\prime }}^{\tau ^{^{\prime \prime
}}}d\tau \frac d{d\tau }\left\{ \sum_{k=1}^rs_k\ln \frac{\Lambda _1^1(\tau
)\cdots \Lambda _k^k(\tau )}{\overline{\Lambda }_1^1(\tau )\cdots
\overline{ \Lambda }_k^k(\tau )}\right\} =
$$
\begin{equation}
\label{c3}=\sum\limits_{k=1}^rl_k\left( Arg{}\Lambda _k(\tau ^{^{\prime
\prime }})-Arg{}\Lambda _k(\tau ^{\prime })\right) ,
\end{equation}

$$
l_k=s_k+s_{k+1}+\cdots +s_r
$$
Now it is a good point to comment on the boundary conditions for the action
functional (\ref{c1}). As we have seen, the equations of motion of the
theory imply the trivial dynamics in the spinning sector (\ref{b11}).
So, in order to allow at least one solution for the variation problem,
the initial and the end points of the trajectory, being
projected onto $\Sigma $, should coincide modulo the equivalence
transformation, i.e.
\begin{equation} Z(\tau ^{\prime \prime })=Z(\tau
^{\prime })\Lambda _0\;,
\end{equation}
for some fixed $\Lambda _0$.
This
leads to the following relation between the boundary values of the gauge
parameters:
\begin{equation} \Lambda (\tau ^{\prime \prime })=\Lambda
_0^{-1}\Lambda (\tau ^{\prime })\Lambda _0
\end{equation}
In particular,
$\Lambda _k^k(\tau ^{\prime })=\Lambda _k^k(\tau ^{\prime \prime })$.
Since $Arg$ is a multivalued function, the difference in rel. (\ref {c3})
under the sum sign equals $2\pi n$, $n\in {\bf Z}$. On the other hand, the
transition amplitudes in quantum theory are obtained by integrating $
\exp (iS)$ over all possible dynamical histories of the particle satisfying
certain boundary conditions. So, requiring the quantum theory to be
invariant under the gauge transformations (\ref{c3}) or, what is the same, $
\exp (iS)$ to be a single valued function, we should restrict $s_1,\ldots
,s_r$ to be integer numbers. Thus we see the quantization rule
for spin in this model.

\section{Quantization}

The identification of the physical phase space of the model with a regular
coadjoint orbit of Poincar\'e group reduces the quantization of the system
to a straightforward exercise in geometric quantization. For the general
theory of geometric quantization, its mathematical and physical contents, we
refer to \cite{Woodhouse}. For the sake of simplicity we will exploit
the rough version of geometric quantization without resorting to the
advanced {\it metaplectic} or {\it half-density quantizations}. The price
of such a simplification is that only the integer spins will gain the
proper description which, however, completely cover the ends of the
present work.  The treatment of half-integer spins will be presented in a
separate paper.

To begin with we identify the set of {\it primary observables} ${\cal A}$ of
the system with the Poincar\'e generators supplemented by constants. After
quantization, the Poisson algebra ${\cal A}$ should be represented by
self-adjoint operators in a Hilbert space ${\cal H}$,
\begin{equation}
{\cal A}\ni f\longmapsto \widehat{f}\in End^{*}{\cal H}
\end{equation}
satisfying Dirac's quantum conditions
\begin{equation}
\label{DC}1\longmapsto id,\quad \widehat{\{f,g\}}=i[\widehat{f},\widehat{g}]
\end{equation}
Since we deal with the elementary quantum system, the Hilbert space ${\cal
H}$ should carry an {\it irreducible} representation of Poincar\'e group.

Now let $B\rightarrow {\cal O}_{m,{\bf s}}$ be an Hermitian line bundle with
connection $\nabla $ and curvature $\Omega $. For such
a line bundle to exist, it is necessary and sufficient to have $\Omega $
satisfying the {\it integrality condition}
\begin{equation}
\label{BZ}\frac 1{2\pi }\int\limits_\sigma \Omega \in {\bf Z\ ,\qquad }
\forall \sigma \in H_2({\cal O}_{m,{\bf s}},{\bf R})\ ,
\end{equation}
$\sigma$ being an arbitrary two-cycle. This is nothing but a geometric
version of the Bohr-Sommerfeld quantization rule in the old quantum theory
(see e.g.  \cite{Woodhouse}, \cite{Simms}). Using the conventional
arguments based on Stokes' theorem, one may verify that eq.(\ref{BZ}) is
always fulfilled when the parameters $ s_1,s_2,...s_r$ entering $\Omega $
are integer numbers as it has been established in previous section.

Let $\Gamma ({\cal O}_{m,{\bf s}})$ denote the space of smooth sections $
\Psi :{\cal O}_{m,{\bf s}}\rightarrow {\bf C}$ of the linear bundle $B$. The
simplest and, in a sense, general way to satisfy the Dirac's conditions (\ref
{DC}) is to assign for every observable $f$ an operator $\widehat{f}$ in $
\Gamma ({\cal O}_{m,{\bf s}})$ defined by the rule
\begin{equation}
\label{q1}\widehat{f}=-i\nabla _{X_f}+f\;,
\end{equation}
where
\begin{equation}
\label{q2}\nabla _{X_f}=X_f-i\vartheta (X_f)\;,
\end{equation}
is the covariant derivative along the Hamiltonian vector field $X_f$
defined by the equation
$$ X_f\rfloor \Omega +df=0 $$
One may endow $\Gamma ({\cal O}_{m,{\bf s}})$ with the Hilbert space
structure by considering square integrable sections w.r.t. the Liouville
measure on ${\cal O}_{m,{\bf s}}$. Then the physical observables become
formally self-adjoint operators. Of course, the space $\Gamma ({\cal
O}_{m, {\bf s}})$ is too wide to represent the Hilbert space of physical
states of the system. The `wave functions', once associated with sections
of $B$, would depend on the whole set of phase space variables $\Gamma $.
As the result, the representation of Poincar\'e group or, what is the
same, the algebra of primary observables ${\cal A}$ is far from
irreducible. That is why, the formulas (\ref{BZ} - \ref{q2}) are referred
to as {\it prequantization}. In order to define the `genuine' wave
functions depending on `half' of the variables one has to `split' $\Gamma
$ into two parts. This is done by choosing {\it polarization} ${\cal P}$,
which may be thought of as a smooth integrable distribution of the
(complex) Lagrange subspaces in complexified tangent bundle $T^{{\bf
C}}({\cal O}_{m,{\bf s}})$. (For the exact definition and detailed
discussion of the concept of polarization see \cite{Woodhouse} .) In the
case at hand, we see that there is a natural Poincar\'e invariant
polarization ${\cal P}$ on ${\cal O}_{m,{\bf s}} $, which is a combination
of the vertical polarization on the cotangent bundle over mass hyperboloid
$p^2+m^2=0$ (momentum representation in the physical literature) and the
positive K\"ahler polarization on ${\cal O}_{ {\bf s}}$. This
polarization can also be viewed as a result of the Hamiltonian reduction
of the polarization generated by the vector fields \{$\partial /\partial
x^A,\partial /\partial \overline{z}_i^B$\}. Now we introduce the subspace
$\Gamma _{{\cal P}}({\cal O}_{m,{\bf s}})\subset \Gamma ({\cal O}
_{m,{\bf s}})$ of {\it polarized sections} extracted by the condition
\begin{equation}
\label{PS}\Psi \in \Gamma _{{\cal P}}({\cal O}_{m,{\bf
s}})\ \Leftrightarrow \ \nabla _X\Psi =0
\end{equation}
for every $X$
tangent to ${\cal P}$. Note that observables from ${\cal A}$ preserve
polarization, i.e. $[{\cal A},{\cal P}]\subset {\cal P}$, and therefore
their action (\ref{q1}) can be consistently restricted to $\Gamma _{{\cal
P}}({\cal O}_{m,{\bf s}})$. On the choice of local trivialization, the
sections (\ref{PS}) are represented by the functions
\begin{equation}
\label{eps}\Psi =\psi (p,Z)e^{-\Phi (Z,\overline{Z})}\ ,
\end{equation}
where $\psi $ are holomorphic in $Z$ and obeying the condition
\begin{equation}
\label{psi}\psi (p,Z\Lambda )=\psi (p,Z)\prod\limits_{i=1}^r(\Lambda
_i^i)^{l_i}
\end{equation}
Together with (\ref{a9}) this condition ensures an independence of physical
states from the $\Lambda $-transformations (the wave functions (\ref{eps})
acquire just a phase factor). It is easy to deduce that general solution to
(\ref{psi}) is given by the homogeneous polynomials of the form \footnote{
Here we use the shorthand notations: $A(l)\equiv A_1\cdots A_l$ and $
z_1^{A(l)}\equiv z_1^{A_1}\cdots z_1^{A_l}$ .}
\begin{equation}
\label{zz}\psi (p,Z)=\psi (p)_{A(l_1)B(l_2)\cdots
C(l_r)}z_1^{A(l_1)}z_2^{B(l_2)}\cdots z_r^{C(l_r)}\ ,
\end{equation}
where the symmetry of the indices is described by the following Young
diagram:

\unitlength=0.63mm \special{em:linewidth 0.4pt}
\linethickness{0.4pt}
\begin{picture}(30.00,40.00)(-50.0,117.0)
\put(10.00,150.00){\line(1,0){64.00}}
\put(74.00,150.00){\line(0,-1){8.00}}
\put(74.00,142.00){\line(-1,0){64.00}}
\put(10.00,142.00){\line(0,1){8.00}}
\put(10.00,142.00){\line(0,-1){8.00}}
\put(10.00,134.00){\line(1,0){50.00}}
\put(60.00,134.00){\line(0,1){8.00}}
\put(10.00,128.00){\line(1,0){39.67}}
\put(49.67,128.00){\line(0,-1){8.00}}
\put(49.67,120.00){\line(-1,0){39.67}}
\put(10.00,120.00){\line(0,1){8.00}}
\put(18.00,150.00){\line(0,-1){16.00}}
\put(66.00,150.00){\line(0,-1){8.00}}
\put(52.00,142.00){\line(0,-1){8.00}}
\put(42.00,128.00){\line(0,-1){8.00}}
\put(18.00,128.00){\line(0,-1){8.00}}
\put(26.00,150.00){\line(0,-1){16.00}}
\put(26.00,128.00){\line(0,-1){8.00}}
\put(14.00,146.00){\makebox(0,0)[cc]{$A_1$}}
\put(14.00,138.00){\makebox(0,0)[cc]{$B_1$}}
\put(14.00,124.00){\makebox(0,0)[cc]{$C_1$}}
\put(22.00,146.00){\makebox(0,0)[cc]{$A_2$}}
\put(22.00,138.00){\makebox(0,0)[cc]{$B_2$}}
\put(22.00,124.00){\makebox(0,0)[cc]{$C_2$}}
\put(46.00,124.00){\makebox(0,0)[cc]{$C_{l_r}$}}
\put(56.00,138.00){\makebox(0,0)[cc]{$B_{l_2}$}}
\put(70.00,146.00){\makebox(0,0)[cc]{$A_{l_1}$}}
\put(34.00,146.00){\makebox(0,0)[cc]{$\cdots$}}
\put(34.00,138.00){\makebox(0,0)[cc]{$\cdots$}}
\put(34.00,124.00){\makebox(0,0)[cc]{$\cdots$}}
\put(34.00,131.00){\makebox(0,0)[cc]{$\cdots$}}
\end{picture}

\noindent
with an explicit skew-symmetry in columns and hidden symmetry in rows.
Besides, in accordance with (\ref{b4}), the coefficients of expansion are
assumed to be $p$-transversal and traceless
\begin{equation}
\label{RE}p^A\psi (p)_{\cdots A\cdots }=0\ ,\qquad \eta ^{AB}\psi
(p)_{\cdots A\cdots B\cdots }=0\
\end{equation}
These coefficients are naturally identified with Fouriau transform of the
ordinary tensor fields on Minkowski space ${\bf R}^{d-1.1}$. Together with
the mass-shell condition
\begin{equation}
(p^2+m^2)\psi (p)_{A(l_1)B(l_2)\cdots C(l_r)}=0\ ,
\end{equation}
eqs.(\ref{RE}) constitute the full set of relativistic wave equations on
irreducible massive tensor fields. The Hilbert space of the particle's
states ${\cal H}_{m,{\bf s}}$ is now associated with space of square
integrable sections (\ref{eps}) (or maybe its completion) w.r.t. the
following Hermitian inner product:
\begin{equation}
\left\langle \Psi _1|\Psi _2\right\rangle =\int\limits_{p^2=-m^2}\frac{d{\bf
p}}{p_0}\int\limits_{{\cal O}_{{\bf s}}}d\mu \overline{\Psi }_1\Psi _2
\end{equation}
where
\begin{equation}
d\mu =d*d\Phi \wedge \cdots \wedge d*d\Phi
\end{equation}
is the Liouville volume form on ${\cal O}_{{\bf s}}$. In principle, one can
perform the integration over ${\cal O}_{{\bf s}}$ explicitly and obtain the
invariant field-theoretical inner product, which is usually used to
normalize the solutions of free field equations
\begin{equation}
\left\langle \Psi _1|\Psi _2\right\rangle =N\int\limits_{p^2=-m^2}\frac{d
{\bf p}}{p_0}\overline{\psi }_{1A(l_1)B(l_2)\cdots C(l_r)}\psi
_2^{A(l_1)B(l_2)\cdots C(l_r)}\ ,
\end{equation}
$N$ being a normalization constant depending on spin.

In conclusion of this section let us note that for a fixed momenta, say, $
\stackrel{\circ }{p}=(m,0,\ldots ,0)$, the quantum states (\ref{zz}) carry
the irreducible representation of $SO(d-1)$. These representations, of
course, can be obtained by the direct quantization of ${\cal O}_{{\bf s}}$.
Hence the above constructions exemplify the general Borel-Weil-Bott theory
establishing a correspondence between the irreducible representations of
compact groups and their integral coadjoint orbits of maximal dimension
\cite{Kirillov}.

\section{The minimal coupling to external fields}

Up to now we have considered the free relativistic particle propagating in
the flat space-time. The intention of this section is to extend action
(\ref{b8}) to the case of particle coupling to an arbitrary background of
gravitational and electromagnetic fields. For these ends we replace the
initial configuration space $R^{d-1,1}\times \Sigma $ of the model with the
Lorentz bundle over the curved space-time $M^{d-1,1}$ with the generic fibre
$\Sigma $. The action functional is generalized to remain invariant under
fiber automorphisms, local $U(1)$-symmetry and general coordinate
transformations on $M^{d-1,1}$. This can be achieved by introducing the
gauge fields associated to the vielbein $e_\mu ^A$, torsion-free spin
connection $\omega _{\mu AB}$, and the one-form of the electromagnetic
potential $A_\mu $. Then the minimal covariantization of (\ref{b8}) yields
\begin{equation}
\label{cc1}
\begin{array}{c}
S=\int
\widetilde{\vartheta }\ , \\  \\
\widetilde{\vartheta }=(p_Ae_\mu ^A-eA_\mu )dx^\mu +*D\Phi
\end{array}
\end{equation}
where $D$ is the Lorentz covariant differential along the particle's world
line
\begin{equation}
Dz_i^A=dz_i^A+dx^\mu \omega _\mu \!^A\!_Bz_i^B\ ,
\end{equation}
and $e$ is an electric charge. Here the mass-shell condition $p_Ap^A+m^2=0$
and rels. (\ref{b4}) are still assumed. The interaction
deforms the original presymplectic structure, but the dimension of its
kernel remains intact: $\dim \ker \Omega =\dim \ker \widetilde{\Omega }=1$.
The last fact means that the number of physical degrees of freedom is the
same as in the free theory and hence the interaction does not give rise to
appearance of unphysical modes. The variation of (\ref{cc1}) leads to the
equations of the form:  $\dot \Gamma =\lambda \widetilde{{\bf V}}\cdot
\Gamma $, where the vector field $ \widetilde{{\bf V}}$ generates the
kernel distribution, more explicitly
\begin{equation}
\label{equ}\stackrel{.}{x}^\mu =\lambda
p^Ae_A^\mu \ ,\qquad \frac{Dp_A}{  d\tau }+eF_{AB}p^B=\frac \lambda
4R_{ABCD}p^BS^{CD}\ ,\qquad \frac{Dz_i^A}{ d\tau }=0
\end{equation}
Here
$S^A\!_B=2i(z_i^A\partial _B^i-\overline{z}_i^A\overline{\partial }
_B^i)\Phi $ is the spinning part of the Lorentz generators, $F_{\mu \nu }$
is a strength tensor of the electromagnetic field and $R_{\alpha \beta
\gamma \delta }$ is a curvature of the space-time. As is seen, the first
equation and the l.h.s. of second one reproduce the motion of a spinless
particle in response to electromagnetic and gravitational fields, whereas
the r.h.s. accounts for the coupling of spin to the geometry of space-time
via the curvature tensor. As for the last equation, it reduces the
dynamics of fiber variables to the parallel transport along the world-line
of the particle. The four-dimensional analog to the equations
(\ref{equ}) was originally derived by K\"unzle \cite{Kunzle} in
somewhat different notations.

\section{Concluding remarks}

Let us summarize. Using the Souriau scheme we have constructed the
mechanical model for the massive spinning particle in $d$-dimensional
Minkowski space. Then the model is generalized to the case of
arbitrary background of electromagnetic and gravitational fields within
the framework of minimal coupling procedure. We have shown that the
geometrical quantization of the free spinning particle gives rise to
massive integer spin irreducible representations of the Poincar\'e group.

A central point in setting up the construction is the use of covariant
realization for the phase space of spin ${\cal O}_{{\bf s}}$ by means of
vector variables subject to certain constraints and equivalence relations.
This enables one to operate in explicitly covariant manner at every stage
of the consideration. In particular, we establish the apparent
relationship between physical wave functions of the theory and Poincar\'e
irreducible tensor fields on Minkowski space.

In order to take into account also the half-integer spin representations,
one should replace, from the very beginning, the proper Lorentz group $
SO_0(d-1,1)$ by its double covering $Spin(d-1,1)$. Then the problem reduces
to the technical question about an appropriate covariant realization for the
${\cal O}_{{\bf s}}$ with the aid of spinor variables. We intend to give a
detailed consideration of this question in the forthcoming publication.

\renewcommand{\thesection}{}

\hspace{-0.5cm}\section{Acknowledgments}

This work is partially supported by the grants Joint INTAS-RFBR 95-829
and RFBR 98-02-16261.

\end{document}